\documentclass[pss,fleqn]{w-art}
\usepackage{times}
\usepackage{w-thm}
\usepackage[]{epsfig,graphicx}
\begin{document}
\DOIsuffix{theDOIsuffix}
\Volume{XX}
\Issue{1}
\Month{01}
\Year{2004}
\pagespan{3}{}
\Receiveddate{2 September 2004}



\title[$\pi$ State Induced by  Impurities]{'$\pi$ State'
Induced by  Impurities with a Repulsive 
Interaction}


\author[G. Litak]{Grzegorz Litak\footnote{Corresponding
     author: e-mail: {\sf g.litak@pollub.pl}, Phone:
+48\,81\,5381573,
     Fax: +48\,81\,5250808} \inst{1,2}} 
\address[\inst{1}]{Department of Mechanics, Technical University of 
Lublin,
Nadbystrzycka 36,
PL-20-618 Lublin, Poland}
\address[\inst{2}]{Max Planck Institute
for the Physics of Complex Systems 
N\"{o}thnitzer Str. 38,
D-01187 Dresden, Germany}
\author[M. Krawiec]{ Mariusz Krawiec \inst{3}}
\address[\inst{3}]{Institute of Physics and Nanotechnology Center,
M. Curie-Sk\l{}odowska University,
 pl. M. Curie-Sk\l{}odowskiej 1, 20-031 Lublin, Poland}
\begin{abstract}
We study the properties of  a quantum impurity embedded
in a superconducting host. The superconductor is described by the negative $U$
Hubbard model while the impurity introduces a repulsive interaction to the system.
We discuss the influence of this `on-impurity' Coulomb repulsion on the local
properties (density of states, electron pairing) of the superconductor. We show the condition of 
$\pi$ - like behaviour, defined as two subsystems having  a phase difference of $\pi$, in the system by 
using a proper   
combination of attractive pairing interaction and repulsive one located at  
impurity site.
\end{abstract}
\maketitle                   





\section{Introduction}
The role of impurities in superconducting hosts has been a subject of intensive
theoretical and experimental studies
\cite{imp_review_Be,imp_review_Ha,imp_review_Li,imp_review_Ta,imp_review_Ka}
recently. Mainly, the bulk properties, like transition temperature or density
of states have been investigated \cite{Bickers}. However in a conventional (as
discussed in this paper) $s$-wave superconductors ($SC$), non-magnetic
impurities have only little effect on the transition temperature according the
Anderson theorem
\cite{Anderson_theorem_An,Anderson_theorem_Mo,Anderson_theorem_Gy}. On the
other hand very little attention has been payed to the understanding the
properties of the local environment of a single impurity, like local density of
states ($DOS$) or spatial variation of the order parameter (see however
\cite{Kim}). The effect of non-magnetic impurities on the local properties of
superconductors with exotic pairing has been studied more extensively in
\cite{Tsuchiura,Vojta} and references therein. The large interest in the
subject of impurities in exotic $SC$ comes from the fact that the Anderson
theorem does not work in the later case and impurities have more drastic
influence on the properties of the high-$T_c$ and other non-$s$-wave
superconductors. In particular, it is widely accepted that high-$T_c$ materials
have $d$-wave pairing state \cite{Annett}, which means that $SC$ order
parameter changes its sign under $\pi/2$ rotation. This leads to the so called
$\pi$-phase behaviour and can be seen in nonuniform systems, like those with
surfaces, vortices, cracks, twin boundaries or impurities
\cite{pi_junction_Ka,pi_junction_Lo}.

A system is regarded as in the $\pi$-phase if there is a sign change of the
order parameter between two subsystems. The simplest example is the junction
made from two superconductors with the phase of the order parameter equal to
$\pi$ \cite{pi_junction_Ka,pi_junction_Lo}. In this case Josephson current
becomes negative in contrast to the usual $0$-phase junction. An other
example are granular high-$T_c$ materials which can likely form network of
microscopic $\pi$-junctions \cite{Sigrist} between small regions with different
phases of the order parameter. In such systems the zero-energy Andreev bound
states, zero-bias conductance peaks, paramagnetic Meisner effect and
spontaneously generated currents take place
\cite{pi_junction_Ka,pi_junction_Lo}.

The situation is very similar to those in which paramagnetic impurities with
strong on-site Coulomb repulsion are placed in classical $s$-wave
superconductor. Due to the proximity effect \cite{Lambert} the $SC$ order
parameter is created on the impurity site. Moreover, because of the fact that
the impurity coupling constant has opposite sign to $SC$ one, there is a sign change
of the order parameter between impurity and surrounded $SC$. Therefore, we regard the system
 as in $\pi$-phase. It is the purpose of the present work to clarify
if such system has all necessary ingredients to regard its as a $\pi$-phase 
state.

\section{The model}

The system is described by negative $U$ Hubbard model \cite{Micnas} with
the Hamiltonian:

\begin{eqnarray}
 H = \sum_{ij\sigma} \left(t_{ij} - \mu \delta_{ij}\right)
     c^+_{i\sigma} c_{j\sigma} +
     \frac{1}{2} \sum_{i\sigma} U_i n_{i\sigma} n_{i-\sigma},
 \label{Hamiltonian}
\end{eqnarray}
where $i$, $j$ label sites of a square lattice, $t_{ij} = -t$ is the hopping
integral between nearest neighbour sites and $\mu$ is the chemical potential.
$U_i < 0$ describes attraction between electrons with opposite spins occupying
the same site $i$. The effect of impurity is introduced in one of the lattice
sites $i_A$ via repulsive interaction $U_{A} > 0$.

In the following we shall work in the Hartree-Fock approximation. To simplify the set
equations we dropped the
Hartree terms ($U_i \langle n_{i\sigma} \rangle$), which means that we have only
off-diagonal impurity induced disorder in the system. So the corresponding
Gorkov equation has the form:

\begin{eqnarray}
 \sum_{j'} \left(
 \begin{array}{cc}   
  (\omega + \mu)\delta_{ij'} - t_{ij'} & \Delta_i \delta_{ij'} \\
  \Delta^{\ast}_i \delta_{ij'} & (\omega - \mu)\delta_{ij'} + t_{ij'}
 \end{array}
 \right)
 \hat G(j',j;\omega) = \delta_{ij}.
 \label{Gorkov}
\end{eqnarray}
In the zero temperature the $SC$ order parameter $\Delta_i$ and the total local
charge $n_i$ are given by self-consistent relations:
%
\begin{eqnarray}
 \Delta_i \equiv U_i \chi_i =
            -  U_i \frac{1}{\pi}\int^{E_f}_{-\infty} d\omega \;
            {\rm Im} G^{12}(i,i;\omega+{\rm i} \epsilon),
 \label{Delta} \\
%
 n_i = -  \frac{2}{\pi}\int^{E_f}_{-\infty} d\omega \;
            {\rm Im} G^{11}(i,i;\omega +{\rm i} \epsilon),
 \label{charge}
\end{eqnarray}
where $E_f$ denotes Fermi energy, and $\epsilon$ is a small positive 
number.
The local quasiparticle density  of states ($LDOS$) can be written 
as follows
\begin{equation}
D_i(\omega)=- \frac{1}{\pi} {\rm Im} G^{11}(i,i;\omega +{\rm i} \epsilon).
\label{density}
\end{equation}

Eqs. (\ref{Gorkov})-(\ref{charge}) have been solved  self-consistently
using one site approximation at two dimensional lattice.
The pairing amplitude $\Delta_A$ has been found at the impurity site.
The surrounding
sites were treated homogeneously as in a bulk (clean) superconductor.

\section{ Results and  discussion}
 
It is well known that the local density of states at an impurity embedded in $SC$
host has bound states within energy gap $\Delta_{bulk}$ which are symmetrically
located with respect to the Fermi energy. Especially this is true if the
on-impurity Coulomb interaction has the same sign as $SC$ one or if it is equal
to zero. In this case bound states always have non-zero energies. However
situation is quite different if the interaction has opposite sign to the $SC$
one. One can show that under special conditions it is possible to get bound
states which are exactly at the Fermi energy.
 
\begin{figure}[h]
\hspace{0cm}
\centerline{
 \epsfig{file=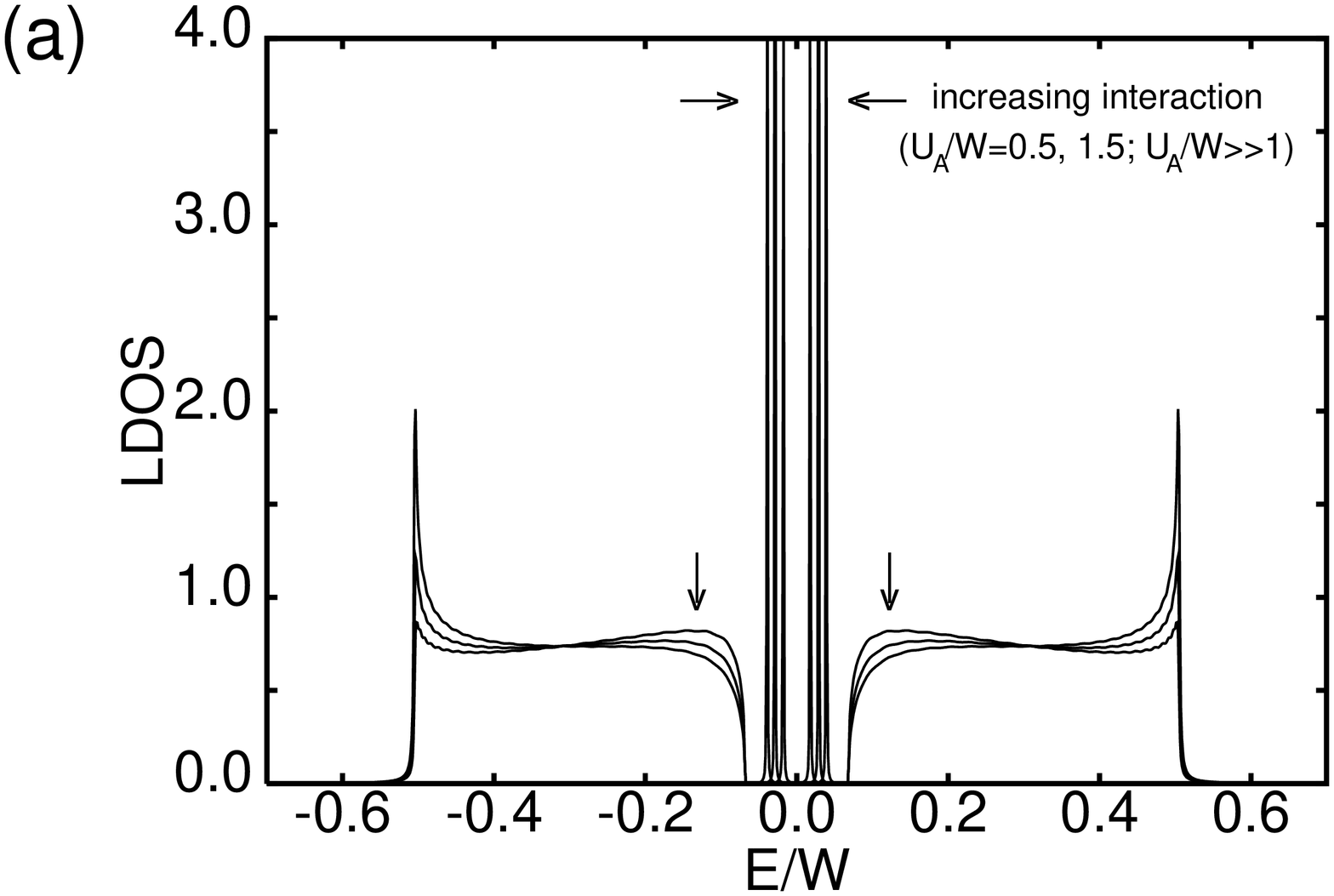,width=7.0cm,angle=-90}}
\centerline{
\epsfig{file=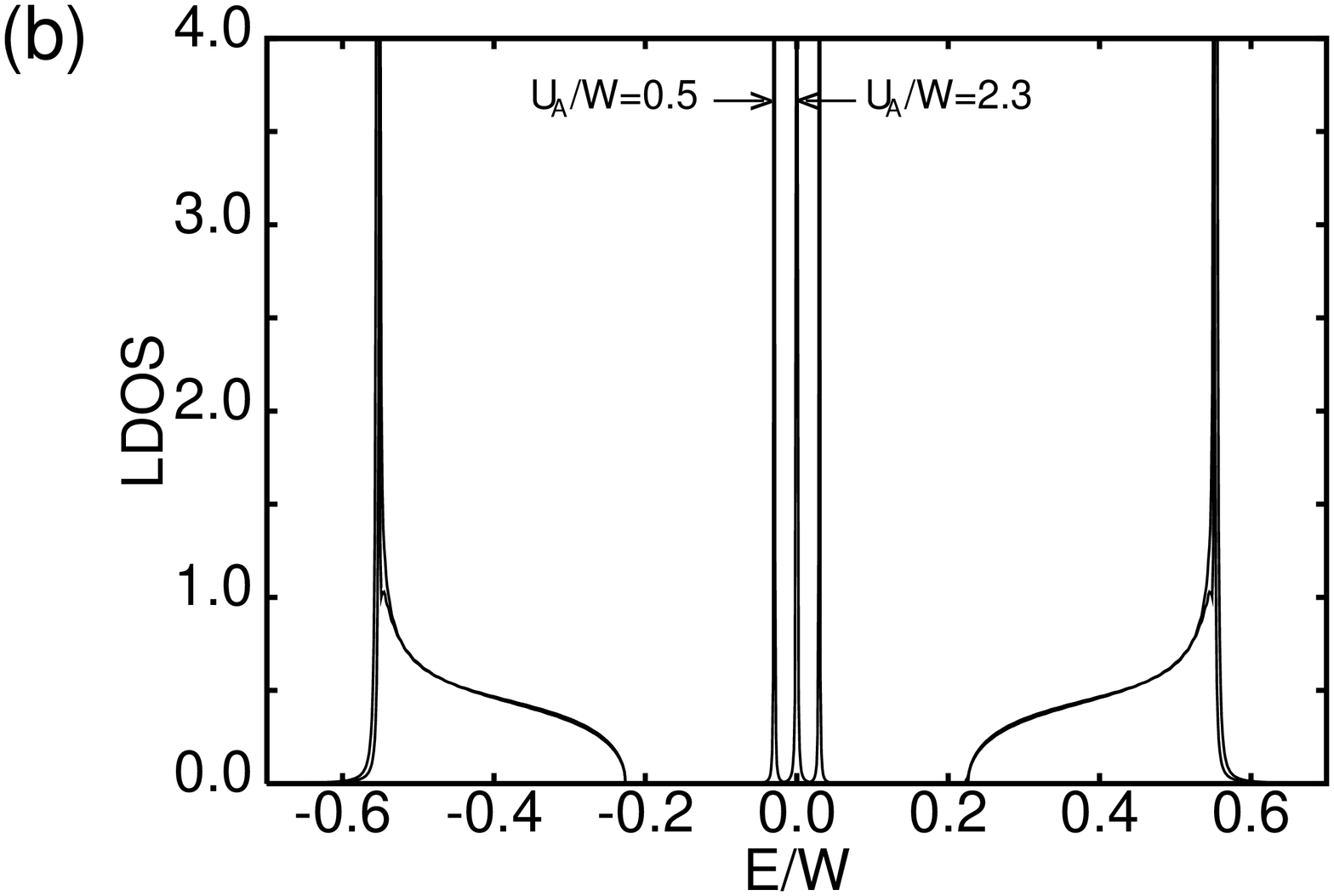,width=7.0cm,angle=-90}}
 \caption{Local density of states for $U_{bulk}/W = -0.3$ (a) and
$U_{bulk}/W =-0.6$ (b). In Fig. 1a vertical and horizontal arrows show the change of $LDOS$ with increasing 
interaction $U_A$.  
}
 \label{fig1}
\end{figure}

\begin{figure}[h]
 \epsfysize=4cm
 \centerline{\epsfig{file=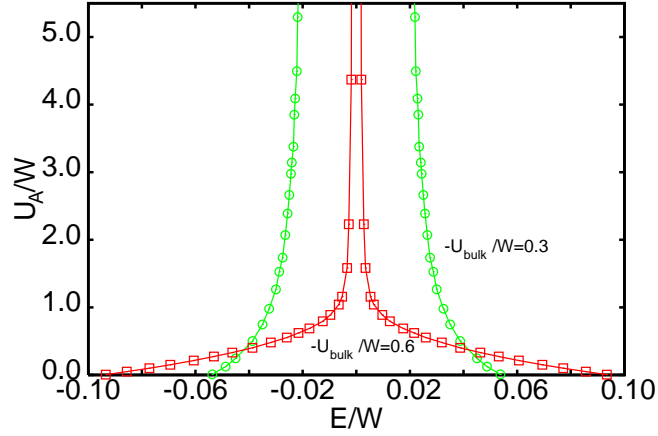,width=7.0cm,angle=-90}}
 \caption{Position of the bound states as a function of $U_A$ for
 $U_{bulk} = -0.3 \; W$ (a) and $U_{bulk} =-0.6 \; W$ (b).}
 \label{fig2}
\end{figure}

The example of the impurity local density of states ($LDOS$) is shown in the
Fig. \ref{fig1} where two situations are depicted.
In Fig. \ref{fig1}a $LDOS$ is plotted for $U_{bulk} = -0.3 \; W$, where $W=8t$ is
a bandwidth, and for a number of on-impurity interactions $U_A > 0$. One can
see that to get zero energy states it is not enough to have $U_A$ of opposite
sign to $U_{bulk}$. In this case bound states have non-zero energies for all
$U_A$. If we increase $U_{bulk}$ (see Fig. \ref{fig1}b) and calculate $LDOS$ one
can find that those bound states do approach zero energy. It can be read from 
Fig. \ref{fig2} where the positions of the bound states are plotted as a
function of $U_A$ for different values of the $U_{bulk}$.

However it is very difficult to judge numerically if those states are really
zero energy states. Therefore we supplemented our study by simple analytical
calculations with constant bulk bare $DOS$. (Note that in numerical
calculations we have chosen $2D$ tight binding $DOS$ with Van Hove singularity
in the middle of the band.) It turns out that the zero energy solution occurs
for only one value of $U_A$ which can be calculated from 
\begin{eqnarray}
\Delta_A \equiv U_A \chi_A = - 2 \pi t^2 \rho(0),
\label{an}  
\end{eqnarray}
where $\rho(0)$ is the bulk bare density of states at the Fermi energy.

\begin{figure}[ht]
\vspace{0.5cm}   
 \epsfysize=4cm
 \centerline{\epsfig{file=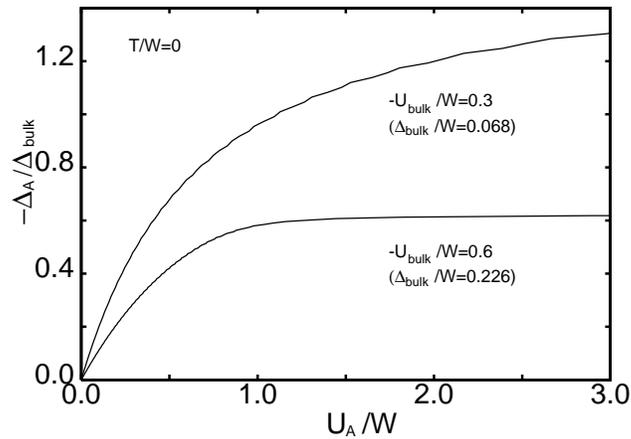,width=7.0cm,angle=-90}}
 \caption{$\Delta_A/\Delta_{bulk}$ plotted as a function of $U_A$. }
 \label{fig3}
\end{figure}

To get such a large value of $\Delta_A$ we have to take large $U_A$ as well as
large value of $\Delta_{bulk}$ because $\chi_A$ in Eq. \ref{an} depends on it
via proximity effect. Physically we can imagine that there could be a strong
on-impurity Coulomb repulsion. However as we can see in the Fig. \ref{fig3} we
cannot take large values of $U_A$ because $\Delta_A$ tends to a constant value as
we increase $U_A$.
The other possibility is to have large values of $\Delta_{bulk}$ which in fact
is impossible to achieve in real materials. We need large values of the bulk 
pairing potential, much larger than the bandwidth.

\section{Conclusions}

 From the present calculations we conclude that it is in general  not possible to get 
states at exactly zero
energy  in the system consisting of a single impurity embedded in $BCS$
superconductor described by the negative $U$ Hubbard model. 
However we were able to find one condition at which such
situation can take place, which seems to be physically impossible to meet it
in reality. 
Note the present calculations were performed assuming 
a rather small scattering rate $\epsilon=10^{-3}t$ (Eq. \ref{density}, Fig. \ref{fig2}).
Of course taking  a large 
enough $\epsilon$ will be
sufficient to broad and eventually connect the two neighbour bound states 
into one at zero energy.

In this paper we have shown results for a half-filled band but we have
also checked  the situation for other fillings and as far as the bound
states are concerned we got no qualitative differences.
 On the other hand present calculations are not fully
self-consistent ones.
We hope that proper treatment of the problem taking
into account the Hartree term in Eq. \ref{Gorkov} and calculating
$\Delta_{bulk}$ at each surrounding site we will be able to give more
precise conditions for $\pi$-like state appearance  in such
systems. 

Note also that relatively large values of the impurity potential $U_A$ (Fig.  \ref{fig1}b)
make Hartee-Fock approximation
questionable  so the
results obtained here should be treated as qualitative.
 Our initial calculations show that fluctuations of the pairing parameter $\Delta_i$ around
the central site would also influence
the critical value of $U_A$ making it even higher. Such an analysis
will be performed systematically
for various band filings $n$.

\begin{acknowledgement}
This work has been partially supported
by the Polish State
Committee for
Scientific Research (KBN), Project No. 2 P03B 06225.
\end{acknowledgement}


\begin{thebibliography}{10}
\bibitem{imp_review_Be}
D. Belitz and T.R. Kirkpatrick {\it Rev. Mod. Phys.} {\bf 66} (1994) 261.
\bibitem{imp_review_Ha}
G. Hara\'n and A.D.S. Nagi, {\it Phys. Rev.} {\bf B 58} (2001) 012503.
\bibitem{imp_review_Li}
G. Litak, {\it Phys. Stat. Sol.} {\bf B 229} (2002) 1427.
\bibitem{imp_review_Ta}
J.L. Tallon {\it et
al.}, {\it Phys. Rev. Lett.} {\bf 79} (1997) 5294.
\bibitem{imp_review_Ka} 
 K. Karpi\'nska
{\it et al.},
{\it Phys. Rev. Lett.} {\bf 84} (2000) 610.

\bibitem{Bickers} N. E. Bickers and G. E. Zwicknagl, {\it Phys. Rev.} {\bf B 36}
(1987) 6746.
\bibitem{Anderson_theorem_An} P. W. Anderson, {\it Phys. Rev. Lett.} {\bf 3}, (1959)
325.
\bibitem{Anderson_theorem_Mo}
                           R. Moradian {\it et al.}, {\it Phys. Rev.} {\bf B 63}
(2001)
024501.
\bibitem{Anderson_theorem_Gy}
                            B. L. Gy\"{o}rffy {\it et al.}, in {\it Fluctuation
Phenomena in
High Critical Temperature Superconducting Ceramics}, eds. M. Ausloos and
A.A. Varlamov
 (Kluwer Academic Publishers NATO ASI Series, Dordrecht 1997) 385.
\bibitem{Kim} H. Kim and P. Muzikar, {\it Phys. Rev.} {\bf B 48} (1993) 3933.

\bibitem{Tsuchiura} H. Tsuchiura {\it et al.}, {\it Journal of the Phys. Soc. of
                    Japan} {\bf 68} (1999) 2510.

\bibitem{Vojta} M. Vojta, R. Bulla, Phys. Rev. {\bf 65}, 014511 (2001).

\bibitem{Annett} J. F. Annett {\it et al.}, in {\it Physical Properties of High
                 Temperature Superconductors} ed. D. M. Ginsberg, World
                 Scientific, Singapore (1996).

\bibitem{pi_junction_Ka} S. Kashiwaya and Y. Tanaka, {\it Rep. Prog. Phys.} {\bf 63}
                         (2000) 1641.
\bibitem{pi_junction_Lo}
                      T. L\"{o}fwander {\it et al.}, {\it Superconduct. Sci.
Technol.}
                      {\bf 14}, (2001) R53.

\bibitem{Sigrist} M. Sigrist and T. M. Rice, {\it Rev. Mod. Phys.} {\bf 67}
                  (1995) 503.

\bibitem{Lambert} C. J. Lambert and R. Raimondi, {\it J. Phys. Condens. Matter}
                  {\bf 10} (1998) 901.

\bibitem{Micnas} R. Micnas {\it et al.}, {\it Rev. Mod. Phys.} {\bf 62} (1991) 113.

\bibitem{recursion} G. Litak {\it et al.}, {\it Physica} {\bf C 251} (1995) 263.
                            
\bibitem{Belzig_thes} W. Belzig, PhD. thesis, Univ. of Karlsruhe (1999).

\bibitem{deGennes} P. G. de Gennes and D. Saint-James, {\it Phys. Lett.} {\bf 4}
(1963) 151.

\bibitem{Hu} C. R. Hu, {\it Phys. Rev. Lett.} {\bf 72}, (1994) 1526.

\bibitem{Belzig} A. L. Fauch\'{e}re {\it et al.}, {\it Phys. Rev. Lett.} {\bf 82}
(1999)
3336.
\end{thebibliography}
\end{document}